\documentclass{ptephy_v1}

\usepackage{color} %
\usepackage{ulem} %

\title{Detection capability of Migdal effect for argon and xenon nuclei with position sensitive gaseous detectors}

\begin{document}

\author[1]{Kiseki~D.~Nakamura}
\author[1]{Kentaro~Miuchi}
\author[2]{Shingo~Kazama}
\author[3]{Yutaro~Shoji}
\author[4,5]{Masahiro~Ibe}
\author[6]{Wakutaka~Nakano}

\affil[1]{Department of Physics, Graduate School of Science, Kobe University, 1-1 Rokkodai-cho, Nada-ku, Kobe, Hyogo, 657-8501, Japan \email{kiseki@phys.sci.kobe-u.ac.jp}}
\affil[3]{Racah Institute of Physics, Hebrew University of Jerusalem, Jerusalem 91904, Israel}
\affil[4]{Institute for Cosmic Ray Research (ICRR), The University of Tokyo, Chiba 277-8583, Japan}
\affil[5]{Kavli Institute for the Physics and Mathematics of the Universe
 (WPI), The University of Tokyo Institutes for Advanced Study,  The
 University of Tokyo, Kashiwa 277-8583, Japan}
\affil[6]{Department of Physics, University of Tokyo, Bunkyo-ku, Tokyo 113-0033, Japan} 

\begin{abstract}

Migdal effect is attracting interests because of the potential to enhance the 
sensitivities of direct dark matter searches 
to the low mass region.
In spite of its great importance,
the Migdal effect has not been experimentally observed yet. 
A realistic experimental approach 
towards the first observation of the Migdal effect
in the neutron scattering
 was studied with Monte Carlo simulations.
In this study, potential background rate was studied together with the event rate of the Migdal effect by a neutron source.
It was found that 
a table-top sized $\sim (30\rm cm )^3$ position-sensitive gaseous detector 
filled with argon or xenon target gas 
can detect characteristic signatures of the Migdal effect
with sufficient rates (O($10^2\sim10^3$) events/day).
A simulation result of a simple experimental set-up showed 
two significant background sources, namely the intrinsic neutrons and the neutron induced gamma-rays.
These background rates were found to be 
much higher than those of the Migdal effect in the neutron scattering.
As a consequence of this study, it is concluded that the experimental observation of the Migdal effect in the neutron scattering
can be realized with a good understanding and reduction of the background.

\end{abstract}
\date{August 2020}

\maketitle

\section{Introduction}
Revealing the nature of the unidentified gravitational source in the universe, or the dark matter, is one of the most important task in particle physics, astroparticle physics and cosmology.
Weakly Interacting Massive Particles (WIMPs) are said to be one of the most attractive candidates and have been searched for decades by various ways~\cite{DM_search_review, DM_search_direct}.
In spite of these intensive searches for the ``standard'' WIMPs with a mass of hundred GeV to several TeV, they have not been discovered yet.
Although the interest and motivation for these standard WIMPs have not decreased~\cite{3TeV_WINO}, some experimental efforts to search for other types of dark matter candidates are being carried out.
Among many types of other dark matter candidates~\cite{ADMX, darkphoton_review}, lighter WIMPs are attracting interests and some experiments have investigated this new frontier~\cite{SuperCDMS, CRESST-III, DarkSide50}.

Low mass WIMPs are conventionally searched for with low threshold WIMPs detectors such as bolometers and ``S2-only analysis'' of two-phase liquid noble gas detectors.
Recently, some groups reported the results and plans for light WIMPs searches using the effect so called the Migdal effect~\cite{XENON1T_migdal, LUX_migdal, EDELWEISS_migdal, CDEX_migdal, SENSEI_migdal, LAr_migdal} %
motivated by 
the recent reformulation of the effect~\cite{migdal_by_ibe} (see also \cite{migdal_by_dolan,Essig:2019xkx}).

The Migdal effect refers to immediate excitation or ionization of the recoil atom at the standard nuclear recoil,
which was first calculated in a different context by A. B. Migdal in 1941 \cite{migdal}.
A sudden change of the nuclear velocity caused by the recoil disturbs the electric potential of the atom. 
When its energy is absorbed by the surrounding electrons, the recoil atom is either excited or ionized.
Then, the excited/ionized atom immediately de-excites through
the X-ray transition, the Auger transition, or the Coster-Kronig
transition, which deposits additional energy inside the detector.
The rate of having the Migdal effect depends on the momentum transfer, but it is very suppressed in the case of light WIMPs.

These Migdal effects can even take place in a very low energy nuclear recoil where the recoil atoms are hardly detectable due to the poor ionization capability. 
Thus the Migdal effect effectively lowers the detection energy threshold of nuclear recoils and drastically improves the sensitivity of the low mass WIMPs searches. In spite of its great importance for the WIMPs searches, the Migdal effect itself has not been observed experimentally yet. A realistic experimental approach, including the potential background sources, towards the first observation of the Migdal effect is discussed in this paper.

\section{Conceptual experimental design}
As introduced in the previous section, the experimental observation of the Migdal effect itself is important for the low mass WIMPs searches. 
We hereby propose a method to detect the characteristic signature of the Migdal effect; a nuclear recoil with an additional electric signal. In order to detect the topology of these two signals, 
position-sensitive gaseous detectors were considered.
Among three possible reactions of the electric signals, we focused on the reaction associated with the emission of the characteristic X-rays so that the electric signal can be spatially-separated from the nuclear recoils with an appropriate choice of the gas and its pressure. 
This ``two-cluster'' event topology can be used 
to select signals out of huge amount of background events to claim a clear observation of the Migdal effect.
Argon gas at 1~atm and xenon gas at 8~atm are chosen for this study so that the characteristic X-ray absorption length would be several cm.

\par

Time projection chambers (TPCs) are widely-used position-sensitive gaseous detectors.
Micro Pattern Gaseous Detectors (MPGD) are well-studied readout devices for gaseous argon TPCs around the normal pressure. A typical three-dimensional spatial resolution of mm or less and an energy resolution of about $30\%$ FWHM at 5.9~keV are realized~\cite{uPIC2003,uTPC2019}. 
A readout system using the  electroluminescence (EL) photons are developed for a high pressure gaseous xenon detector~\cite{AXEL2020}.
The energy resolution of $4\%$ at 30~keV has been demonstrated with the EL signal read by photon detectors with a pitch of 10~mm~\cite{AXEL2020}.
Considering these realistic detectors, 
we assume a detector with detection 
volume of $(30\,\rm cm)^3$ with MPGD and EL readouts for the argon and xenon TPCs, respectively.
Total numbers of the target nuclei for the argon and the xenon detectors are shown in Table~\ref{tab:migdal}.

There are various neutron sources potentially available for this experiment. In this study a continuous neutron beam with an energy of 565~keV and a flux of $1000\,\rm /cm^{2}/sec$ at 1~m is assumed. This is a typical value for a $\rm ^7Li(p,n)^7Be$ reaction 
at an irradiation facility in National Institute of Advanced Industrial Science and Technology (AIST), Japan. 
The cross sections of elastic scattering between 565~keV neutrons and the target nuclei are shown in Table~\ref{tab:migdal}.
It should be noted that, in general, the higher the neutron energy is, the more gamma-ray background rate from the detector and the laboratory materials is expected. Thus we choose this energy as a typical neutron energy instead of more common sources like DT (deutron-triton) or DD (deutron-deutron) generators. 

\section{Signal}\label{sec:signal}
In this section, Migdal effect signals are discussed.
First, an overview of the features of the Migdal effect including the expected event rate is given based on Ref.~\cite{migdal_by_ibe}.
Then a more realistic MC simulation study on the detection possibility of the characteristic features of the Migdal effect is described.

\subsection{Signal Overview}
\begin{figure}[htbp]
	\centering
	\includegraphics[width=0.7\linewidth]{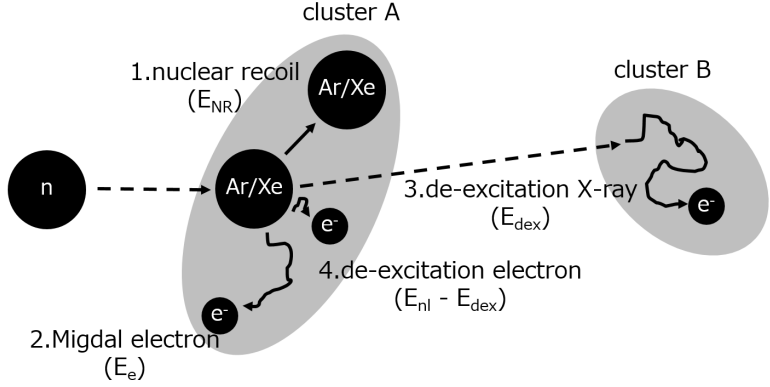} 
	\caption{Shcematis mechanism of the reactions related to the Migdal effect.} 
	\label{fig:event_topology} 
\end{figure}

Probabilities of the Migdal effects depends on the quantum numbers $(n,l)$ of the associated electron.
When an electron with a large principal quantum number ($n$) is involved, the probability of the Migdal effect is large but the additional electric energy is small because of the small binding energy.
Observations of the Migdal effects with large principal quantum numbers, especially with $n\geq 3$ for xenon, are important for further searches and possible discoveries of the WIMPs~\cite{XENON1T_migdal}.
However, the electric energies of these reactions are typically less than 1~keV and 
existing detectors are not suitable for a clear observation with a two-cluster topology. 
Therefore, %
$(n,l)=(1,0)$ cases (1s, K shell) for both argon and xenon where additional electric energies can be detected with practical detectors
are chosen for the first observation of the Migdal effect.
The branching ratios of the Migdal effect for the $(n,l)=(1,0)$ cases are listed in Table~\ref{tab:migdal}.
Figure~\ref{fig:event_topology} illustrates the reactions we expect to observe. 
The first reaction is the nuclear recoil (1) with a recoil energy of $E_{\rm NR}$ and the Migdal effect (2).
The orbital electron that receives energy by the Migdal effect transitions to another orbit or goes out of the atom.
Since the probability of latter process is several orders of magnitude higher than the former one, the latter process is studies in this paper.
The recoil nuclei and the Migdal electron are detected as one cluster (cluster A). 
A de-excitation takes place following the emission of the Migdal electron.
The de-excitation can be seen as a characteristic X-ray (3) or an Auger electron.
Probabilities of de-excitation of K-shell hole by X-ray emission, or the  fluorescence yield, are listed in Table~\ref{tab:migdal}. Characteristic X-ray can be detected at a distance of several cm with the detector condition we assume so that a  second cluster (cluster B) with a fixed energy ($E_{\rm dex}$) is expected.
This topology information will provide us with a clear signal of the Migdal effect together with a powerful background rejection.
Finally, multiple Auger electrons and de-excitation X-rays with a total energy of $E_{\rm nl}-E_{\rm dex}$ are emitted for the energy conservation, where $E_{\rm nl}$ is the binding energy of the $(n,l)$ electron (4).

At the leading order, the probabilities of the Migdal effect
are known to be proportional to the square of the momentum transfer to the Migdal electron $q_{\rm e}$.
Probabilities for a given $q_{\rm e}$ are calculated by scaling 
the values shown in Table 2 of Ref.~\cite{migdal_by_ibe}  
with $(q_{\rm e}/511\,\rm eV)^2$.
$q_{\rm e}$ is calculated
as follows,
\begin{equation}
    q_{\rm e}^2=\frac{2m_{\rm e}^2E_{\rm NR}}{m_{\rm N}},
\end{equation}
where $m_{\rm e}$ is the electron mass and $m_{\rm N}$ is the target nuclear mass.
Table~\ref{tab:migdal} shows the scaling factors for maximum recoil energies  $E_{\rm NR}^{\rm max}$ by an irradiation with 565~keV neutrons. 
Here $E_{\rm NR}^{\rm max}$ is known by 
\begin{equation}
E_{\rm NR}^{\rm max}=\frac{4m_{\rm n}m_{\rm N}}{(m_{\rm n}+m_{\rm N})^2}E_{\rm n}.
\end{equation}
Here $m_{\rm n}$ is the neutron mass and $E_{\rm n}$ is the neutron energy. 
It is seen that the scaling factor for argon is one order of magnitude larger than that for xenon 
because $m_{\rm N}$ is smaller and $E_{\rm NR}$ is consequently larger for a given energy of neutrons.
Expected event rates calculated 
for the experimental and physical conditions discussed so far are shown in the final row of Table~\ref{tab:migdal}. Here the event rate for the nuclear recoils associated with characteristic X-rays from the Migdal effect are shown.
The rates ($O(10^2\sim10^3)$ events/day) themselves without the consideration of any backgrounds are encouraging ones. 
In reality, the background rates without any reduction are much larger than these signal rates. It is one of the important points  
of this study to discuss a realistic method to discriminate background events with the event topologies, 
which will be discussed in Section~\ref{sec:bg}.

It should be emphasized that our calculation and discussed measurement are only for the isolated atoms. For the application to the dark matter searches, it is also important to test the Migdal effect in the liquid medium.

\begin{table}
    \centering
    \begin{tabular}{c|c|c}
        \hline
        target gas & Ar 1 atm $(30\,\rm{cm})^3$ & Xe 8 atm $(30\,\rm{cm})^3$ \\
         \hline
        number of nuclei & $7.26\times10^{23}$ & 5.81$\times10^{24}$ \\
        cross section for 565 keV neutron & 0.65 barn & 6.0 barn \\
        Migdal branching & $7.2\times10^{-5}$ & $4.6\times10^{-6}$ \\
        fluorescence yield (K shell) & 0.14 & 0.89 \\
        scaling factor $(q_{\rm e}^{\rm max}/511\,\rm eV)^2$ & 2.92 & 0.280 \\
        event rate & 603 events/day & 975 events/day \\
         \hline
    \end{tabular}
    \caption{Typical values of parameters for estimating the Migdal effect. The branching ratios for $(n,l)=\rm 1s$ and $q_{\rm e}=511\,{\rm eV}$ are shown.  }
    \label{tab:migdal}
\end{table}

\subsection{Signal Simulation}
A simple calculation showed an encouraging signal event rate as an ideal case.
A more realistic Geant4~\cite{Geant4} Monte Carlo (MC) simulation study was then performed.
Four particles, listed as (1)$\sim$(4) in the followings, are generated for the MC simulation.

\begin{enumerate}
  \item A recoil nucleus with a recoil energy of $E_{\rm NR}$ following the probability distribution discussed in Ref.~\cite{migdal_by_ibe} is generated. The position is homogeneously distributed in the detection volume. 
  Here $E_{\rm NR}$, recoil angle $\theta$ and the transition energy to the electron ($\Delta E$) have correlations as shown in equation (4.18) in Ref.~\cite{migdal_by_ibe}.
  \item A Migdal electron with an energy $E_{\rm e}=\Delta E - E_{\rm nl}$ is generated. 
  \item A de-excitation X-ray with an energy $E_{\rm dex}$ is generated. %
  K$_\alpha$ or K$_\beta$ X-ray with known intensity ratio is generated with its characteristic energy.
  \item De-excitation electron with an energy $E_{\rm nl}-E_{\rm dex}$ is generated. 
\end{enumerate}

  The initial directions of the particle generated in (2)-(4) are isotropic.

\begin{figure}[htbp]
	\centering
	\includegraphics[width=0.99\linewidth]{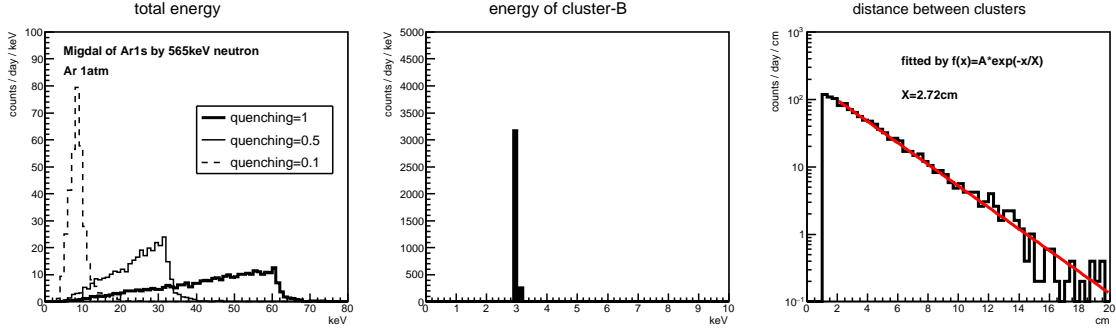} 
	\caption{Signal MC simulation results with argon gas. 
The spectra of the total energy, energy spectra of cluster B, and the distributions of the distance between the two clusters are shown in the left, center, and right, respectively.} 
	\label{fig:Ar_migdal} 
\end{figure}

\begin{figure}[htbp]
	\centering
	\includegraphics[width=0.99\linewidth]{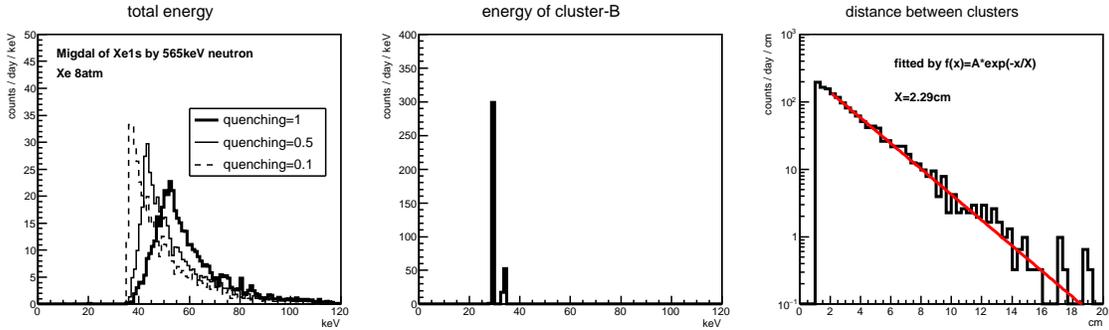} 
	\caption{Signal MC simulation results with xenon gas. 
The spectra of the total energy, energy spectra of cluster B, and the distributions of the distance between two clusters are shown in the left, center, and right, respectively.}
	\label{fig:Xe_migdal} 
\end{figure}

Figure~\ref{fig:Ar_migdal} and \ref{fig:Xe_migdal} show the results of the signal simulations for argon and xenon targets, respectively.
The total energy spectra, energy spectra of the X-rays (cluster B), and the distributions of the distance between two clusters are shown in the left, center, and right panels, respectively.
Here the events with any energy deposition within 1~cm from the wall were rejected (fiducial cut) in order to reject the background charged particles  from the wall of the detector.
Another criteria for the event selection was to require the distance between two clusters to be larger than 1~cm.

The left plots show the quenching factor dependence of the total energy. 
Three cases with the quenching factors of 1, 0.5, and 0.1 are shown. The case of the quenching=1, where the recoil nuclei would show the same ionization efficiency with electrons, is shown as a reference.
Since the quenching factor depends on the detector parameters such as the density of the medium, impurities, and 
the electric field, it needs to be measured at an actual detector condition.
The real quenching factors can be between 0.1 and  0.5, 
and thus energy spectra for these two cases 
are shown as references. 
The real spectrum would be expected somewhere between these two spectra.
The effect of quenching appears more clearly in the spectrum with argon target. This is because the nuclear recoil energy is larger than the energy of the characteristic X-ray. 
Hereafter we will discuss with the case of quenching=1, {\it i.e.} without considering the quenching. This would help to understand the real recoil energy and can be certified since the event selections can be tuned with a measured quenching factor prior to the Migdal effect experiments. 

In the center plots, it is seen that the energy spectra of cluster B make peaks, so that the background can be efficiently rejected by requiring two clusters with either one having the energy of the %
characteristic X-ray. It should be noted that the quenching factors need not to be considered for the X-ray signals. 

The distance distributions between two clusters shown in the right plots have exponential shape representing the absorption length of X-rays.
The best-fit curves shown with red lines are consistent with the ones for the expected absorption length (2.95~cm and 2.19~cm for argon and xenon, respectively).
This shape would be a very good evidence of the Migdal effect in contrast to the flat distribution for the neutron multiple scatterings or square to the distance for accidental backgrounds.

\section{Background}\label{sec:bg}
Background studies were carried out with Geant4 MC simulations.
In this study, it turned out that there exist two significant background sources,  namely the intrinsic neutrons and the neutron induced gamma-rays.  
An unavoidable intrinsic neutron background caused by the incident neutron on the target gas will be discussed first.
Then, a gamma-ray background from the $({\rm n}, \gamma)$ reaction in the chamber and the laboratory material will be evaluated.

\subsection{Intrinsic neutron background}
Nuclear recoils by the incident neutrons associated with another energy deposition 
will become 
unavoidable background events 
because their event topology would look like the Migdal signal events discussed in the previous section. 

First, accidental coincidence background where two independent nuclear recoils take places within a time resolution of the TPC ($\sim 10~\mu\rm{s}$) is studied.
The probabilities of accidental coincidence are $O(10^{-4})$ and $O(10^{-2})$ for the argon and xenon targets, respectively.
Since the cluster distance of accidental coincidence event follow quadratic function, a further $O(10^{-2})$ reduction is expected.
For the argon case, the event rate is smaller than that of the Migdal signal.
For the xenon case, the accidental coincidence can be rejected by requiring the energy of cluster~B to be 30~keV
since the maximum recoil energy is about 18~keV.
We then studied the background events created by interactions of neutrons and target nuclei, mainly multiple scatterings and inelastic scatterings, as an intrinsic background.
The key parameters to reject these intrinsic backgrounds are the energy of the cluster~B and the distance between two clusters as shown in the center and right panels of Figure~\ref{fig:Ar_migdal} and Figure~\ref{fig:Xe_migdal}.
The rate and the rejection efficiency of the background events were studied with the MC simulations.
In order to study the pure effect of this intrinsic background, 
a detector with a gas-only target of $(30\,\rm cm)^3$ was prepared and irradiated with neutrons of the same energy (565~keV) as the signal study.
\par
\begin{figure}[htbp]
    \begin{minipage}{0.49\hsize}
	    \centering
	    \includegraphics[width=0.99\linewidth]{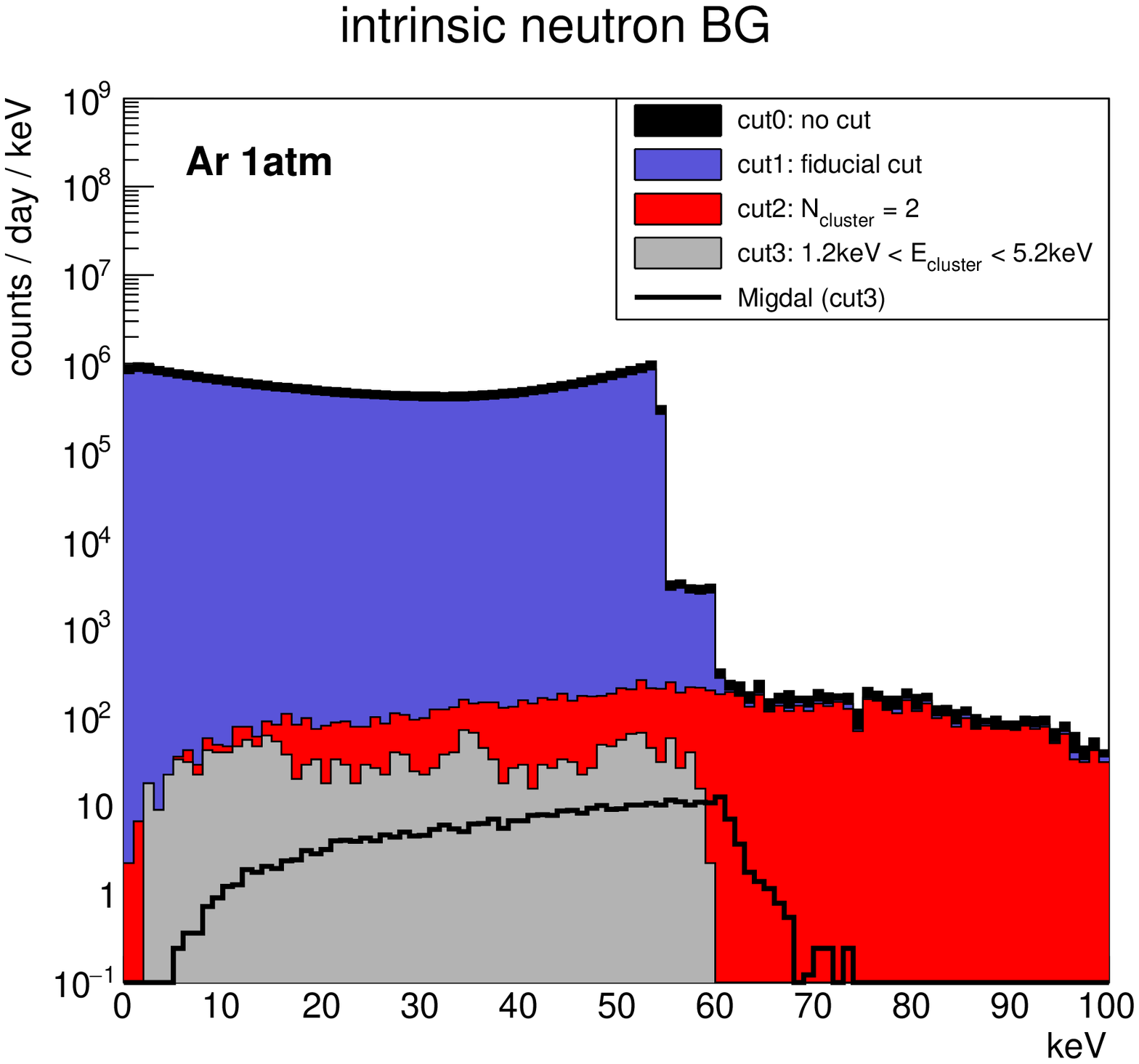} 
	    \caption{Simulated total energy spectra of the intrinsic neutron background events for the argon target. Black-filled histogram is the raw energy spectrum. Blue, red, and gray ones are those after cut 1, 2, and 3, respectively. The black-solid line is the energy spectrum of the Migdal effect.} 
	    \label{fig:Ar_neutron} 
    \end{minipage}
    \begin{minipage}{0.49\hsize}
	    \centering
	    \includegraphics[width=0.99\linewidth]{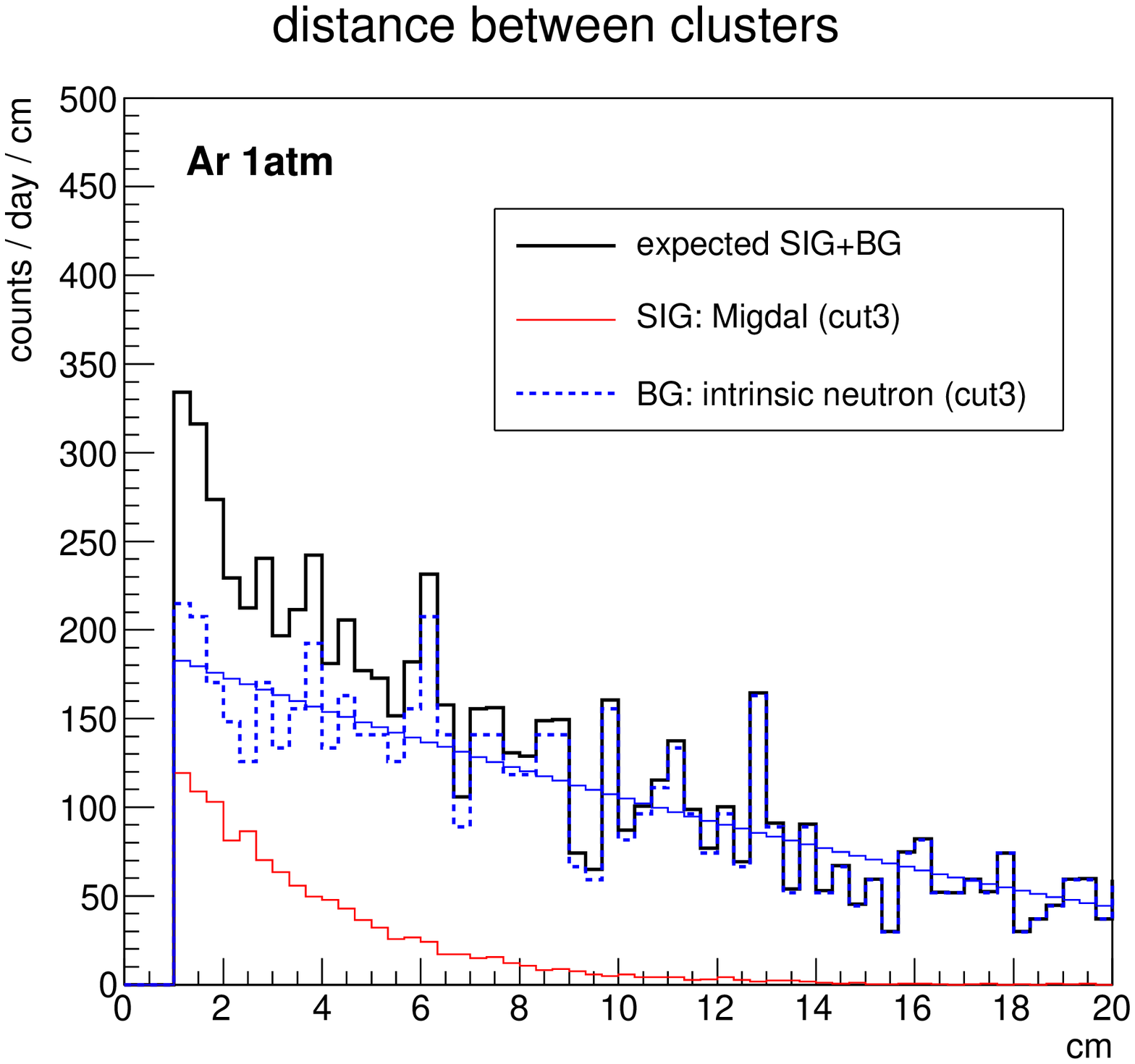} 
	    \caption{MC simulation results on the distance between two clusters for the argon target. Red, blue, and black histogram show the intrinsic neutron background events, signal events and sum of these two components.} 
    	\label{fig:Ar_neutron_length} 
    \end{minipage}
\end{figure}

Figure~\ref{fig:Ar_neutron} shows the total energy spectra obtained by the argon-target MC simulations.
The spectrum without any cut is shown with the black-filled histgram (cut 0).
The spectrum after the fiducial cut, same as the signal selection, is shown with the blue-filled histogram (cut 1).
Elastic scatterings make dominant contribution below 60~keV. 
The background rate is found to decrease by about 4 orders of magnitude
by requiring two clusters (cut 2, red-filled histogram). 
The background rate above 60~keV is further decreased by 
requiring the presence of a cluster with an energy of characteristic X-ray (3.2$\pm$2~keV, cut~3).
This is one of the advantages of using a position-sensitive gaseous detector. 
The spectrum of the final background sample is shown with the grey-filled histogram.
The energy spectrum of the Migdal events is shown with a black line for comparison.
The background rate was found to be a few times larger than the signal rate with a similar energy spectrum.
It should be noted that the energy resolution is not taken into account, which would make the discrimination of these two spectra difficult. 
The possibility of extracting the Migdal events out of the background would be the use of the distance of two clusters which is shown in 
Figure~\ref{fig:Ar_neutron_length}. Here the spectra of the background (cut 3), the signal, and the sum of these two components are shown with red, blue, and black lines, respectively. 
The remaining backgrounds of the intrinsic neutron are neutron multiple scattering events.
The mean free path of the neutron and the argon nucleus is 5.7 $\times 10^4$~cm and the shape of the distribution is determined by the detector size.
The background distance distribution (blue) is in good agreement with the efficiency curve of detecting two clusters in a $(28~\rm cm)^3$ volume.
The cluster-distance distribution of the Migdal effect events is found to be more steep, showing a shorter absorption length of the characteristic X-ray.
Therefore, the observation of the Migdal effect can be claimed by extracting the signal rate from the cluster-distance distribution.

\par
\begin{figure}[htbp]
    \begin{minipage}{0.49\hsize}
	    \centering
	    \includegraphics[width=0.99\linewidth]{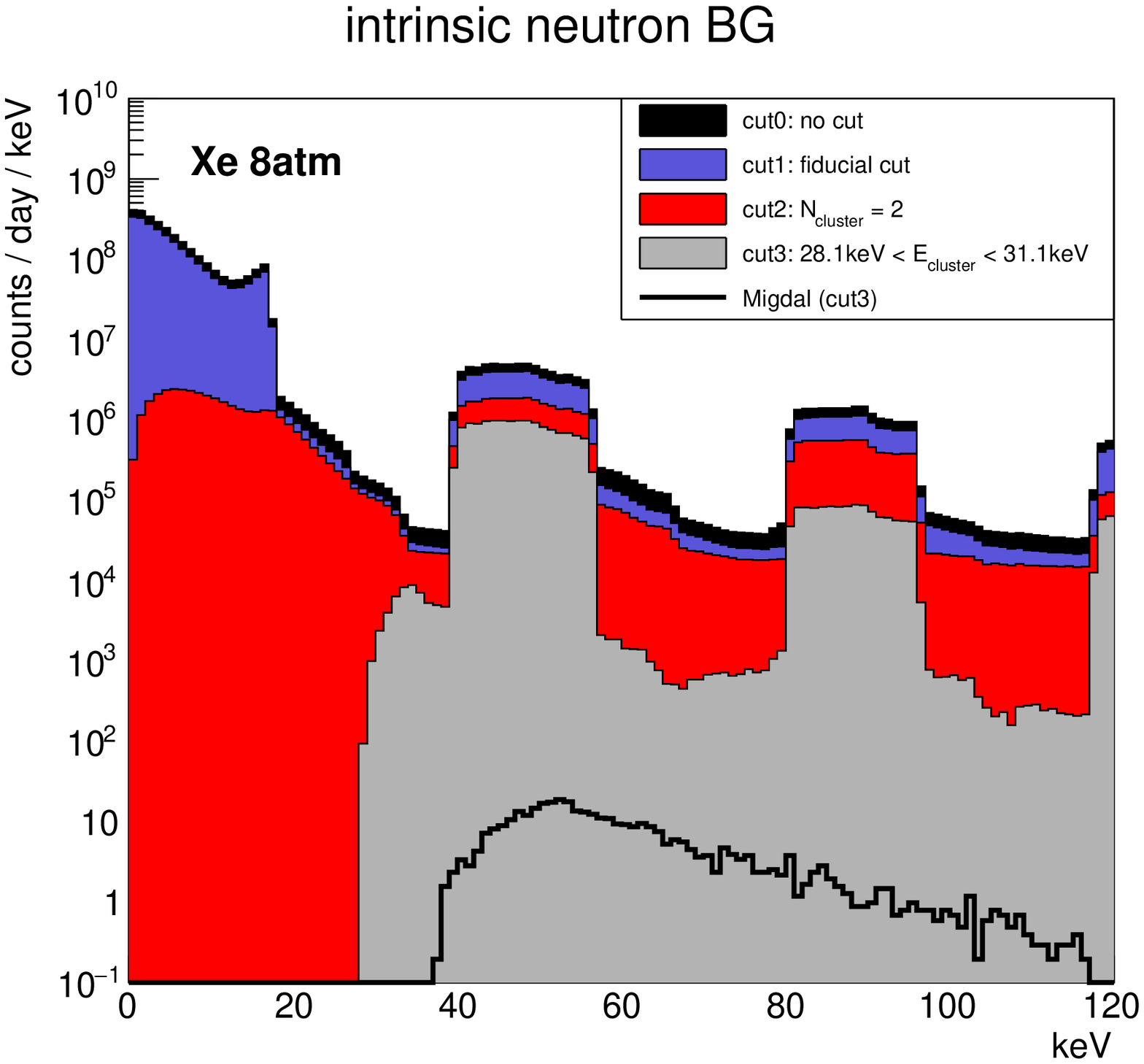} 
	    \caption{Simulated total energy spectra of the intrinsic neutron background events for the xenon target. Black-filled histogram is the raw energy spectrum. Blue, red, and gray ones are those after cut 1, 2, and 3, respectively. The black-solid line is the energy spectrum of the Migdal effect.} 
	    \label{fig:Xe_neutron} 
    \end{minipage}
    \begin{minipage}{0.49\hsize}
	    \centering
	    \includegraphics[width=0.99\linewidth]{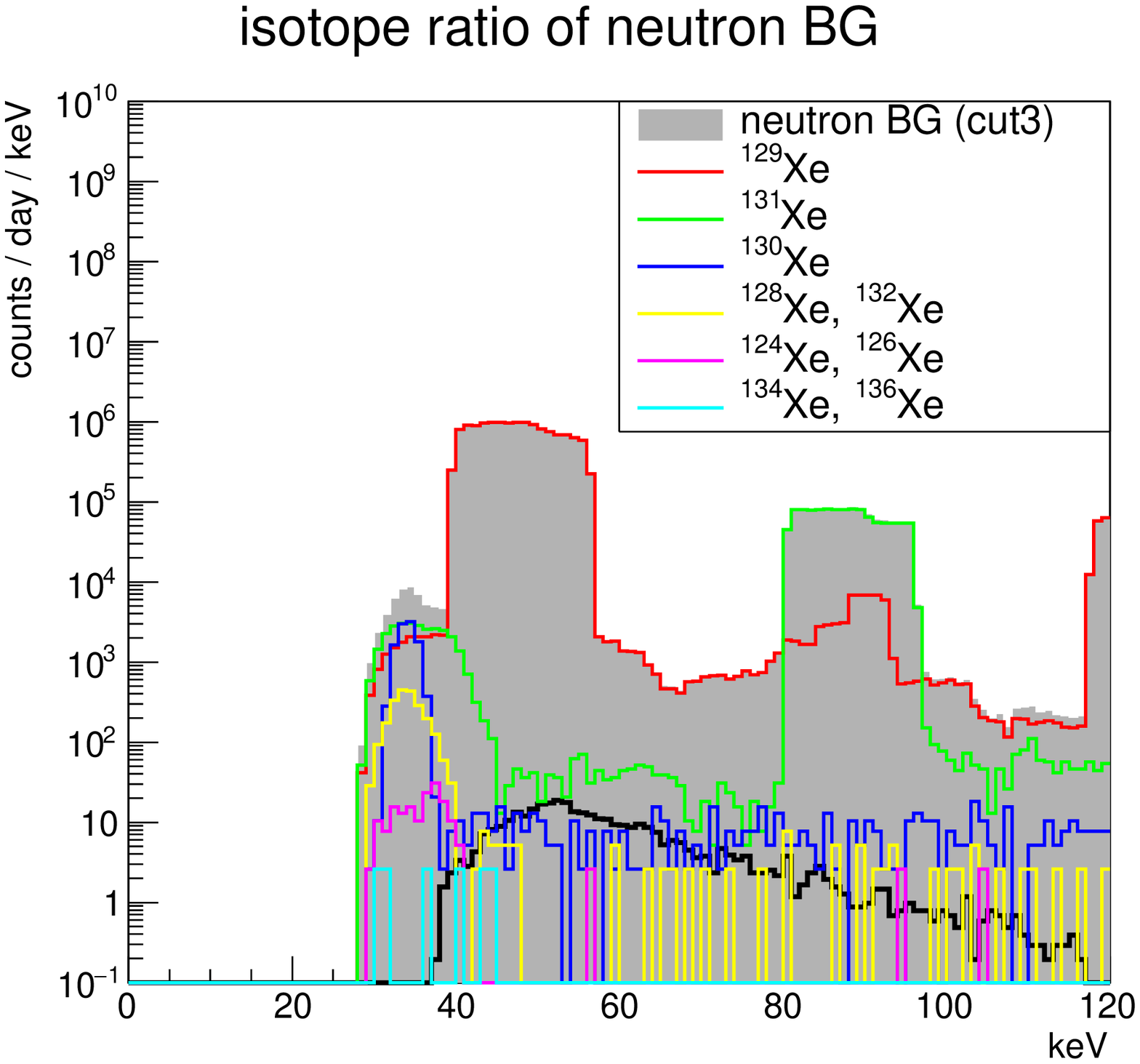}
	    \caption{Isotope-breakdown of the intrinsic neutron background events for the xenon target.} 
	    \label{fig:Xe_neutron_isotope} 
    \end{minipage}
\end{figure}

Figure~\ref{fig:Xe_neutron} shows the total energy spectra obtained by the xenon-target MC simulations.
Cut criteria 0 to 2 are same as the ones for the argon-target case. 
Cut 3 requires that one of the clusters has the energy of the K$\alpha$ X-rays of xenon, 29.6~keV $\pm$ 1.5~keV. 
Here an realistic energy resolution of about 5\% FWHM was assumed.
The count rate of the Migdal effect is shown with a solid black line. 
The result shows a much higher background rate than that of the argon-target case. The reason of higher background rate can be explained by Figure~\ref{fig:Xe_neutron_isotope} which shows the isotope-breakdown of the background energy spectrum.  It is seen that the peak around 40-60~keV and the continuum component over 60~keV are due to $\rm {}^{129}Xe$ (natural abundance 26.4~$\%$) which has an inelastic scattering associated with 40~keV or 120~keV $\gamma$-rays. These inelastic scatterings make two clusters, one by the nuclear recoil and another by the X-ray, and make the discrimination difficult even with the event topologies. Next-largest contribution is $\rm {}^{131}Xe$ (natural abundance 21.2~$\%$) which has an emission of 80~keV $\gamma$-rays. Although the contribution of $\rm {}^{131}Xe$ is much smaller than that of $\rm {}^{129}Xe$, it is still larger than the count rate of the Migdal effect. The third one comes from $\rm {}^{130}Xe$ (natural abundance 4.1~$\%$) which is associated with high energy ($\sim$MeV) gamma-rays. Partial energy depositions of Compton-scattered electrons from these gamma-rays make the continuum component of the background spectrum. Other isotopes make background less than the Migdal effect. Among the various isotopes, it is seen that $\rm {}^{134}Xe$ and $\rm {}^{136}Xe$ which have natural abundances of  10.4~$\%$ and 8.9~$\%$ makes less background than the Migdal effect. 
In this MC study, the background contribution of $\rm {}^{133}Xe$ which is produced by the neutron capture of $\rm {}^{132}Xe$ and emits 81~keV prompt gamma-rays is not taken account because of a long half-life of 5.2 days.
Since the rate is estimated to be around $O(10^4)$ counts/day, it is necessary to apply energy selection of $<$80~keV or remove $\rm {}^{132}Xe$ by isotope separation.

The possibility of the Migdal effect observation with xenon gas becomes realistic with an isotope enrichment of the isotopes larger than 
A=133. More than four orders of reduction for $\rm {}^{129}Xe$
and $\rm {}^{131}Xe$ are required considering the original contributions of these isotopes to the background spectrum. 
With an additional energy selection between 60~keV and 80~keV, the Migdal effect can be observed with xenon gas.

\subsection{Gamma-ray background}
\begin{figure}[htbp]
    \begin{minipage}{0.49\hsize}
	    \centering
	    \includegraphics[width=0.99\linewidth]{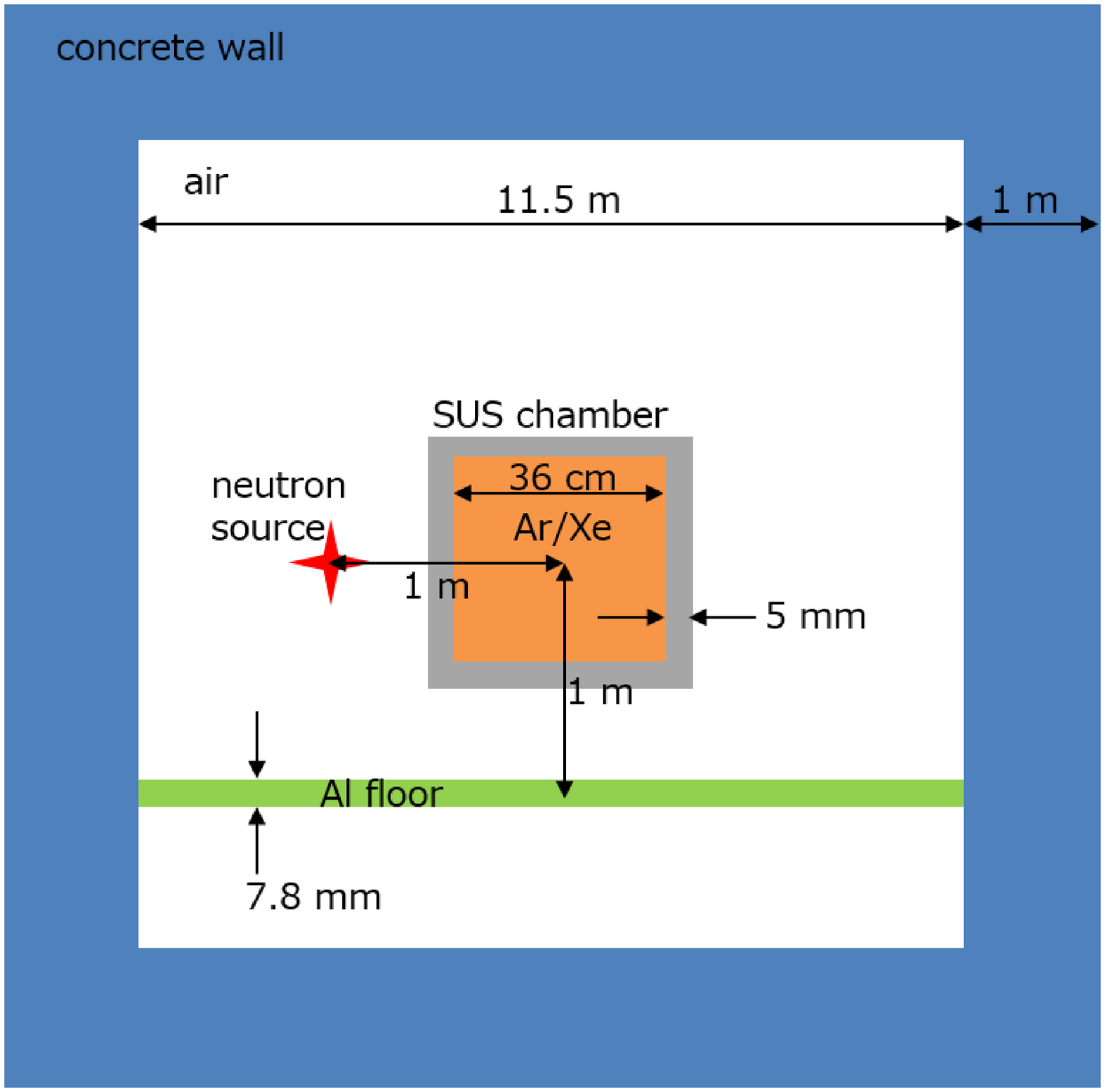} 
	    \caption{A geometry for the gamma-ray background MC simulation.
	    The figure is not to scale.} 
	    \label{fig:BGMCsetup} 
    \end{minipage}
    \begin{minipage}{0.49\hsize}
	    \centering
	    \includegraphics[width=0.99\linewidth]{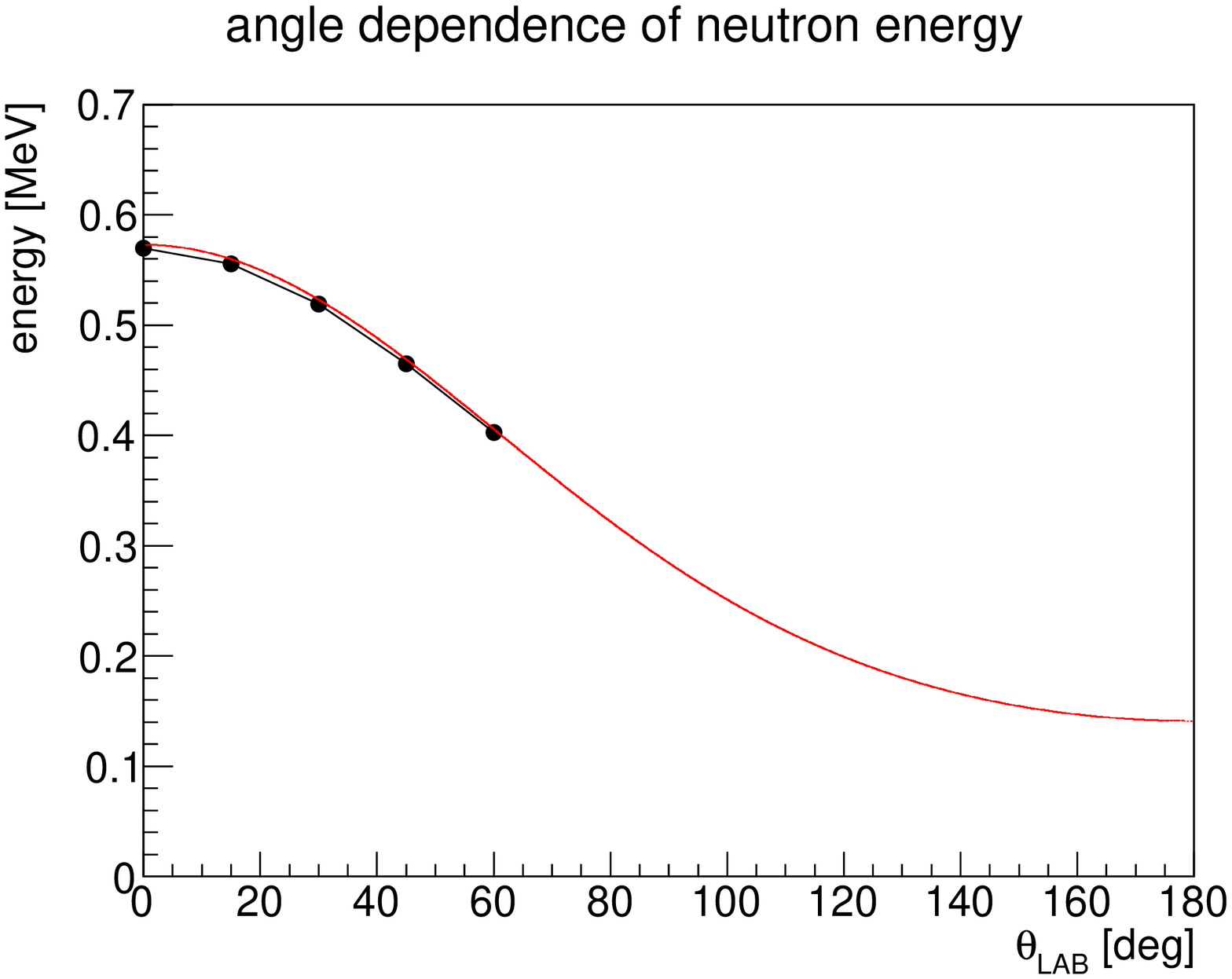}
	    \caption{Angular dependence of neutron energy due to $\rm ^7Li(p,n)^7Be$ reaction.
	    Black dots are simulation results by AIST, red lines are calculation results from kinematics.
	    The neutrons used to estimate the gamma ray background due to the $(\rm n,\gamma)$ reaction with the structure were generated according to the red line.}
	    \label{fig:neutron_energy_angle} 
    \end{minipage}
\end{figure}

Neutrons interact with materials of the laboratory and the detector
and generate gamma-rays via (n,$\gamma$) reactions.
The production rates and the energies of gamma-ray backgrounds depend on the laboratory geometry and detector design.
Here, typical gamma-ray backgrounds for a simplified geometry shown in Fig.~\ref{fig:BGMCsetup} are studied.
A stainless chamber with a thickness of 5~mm
is implemented enclosing a gas medium with a volume of $(36\,\rm cm)^3$.
The detection volume is kept same as the one used in the signal simulation $(30\,\rm cm)^3$ while a 3~cm buffer is prepared for each plane.
The neutron source is set at a position 1~m away from the detector center, same condition as the signal simulation. 
For the gamma-rays from the laboratory materials, 
neutrons from the (p,Li) reaction are generated to directions taking account of the energy and rate dependence. The energy-angle correlation is show in Fig.~\ref{fig:neutron_energy_angle}. 
The detector is set at the center of a laboratory with a size of $(11.5\,\rm m)^3$. An aluminum 
floor with an effective thickness of 7.8~mm 
is constructed 1~m below the detector center.

\begin{figure}[htbp]
    \begin{minipage}{0.49\hsize}
	    \centering
	    \includegraphics[width=0.99\linewidth]{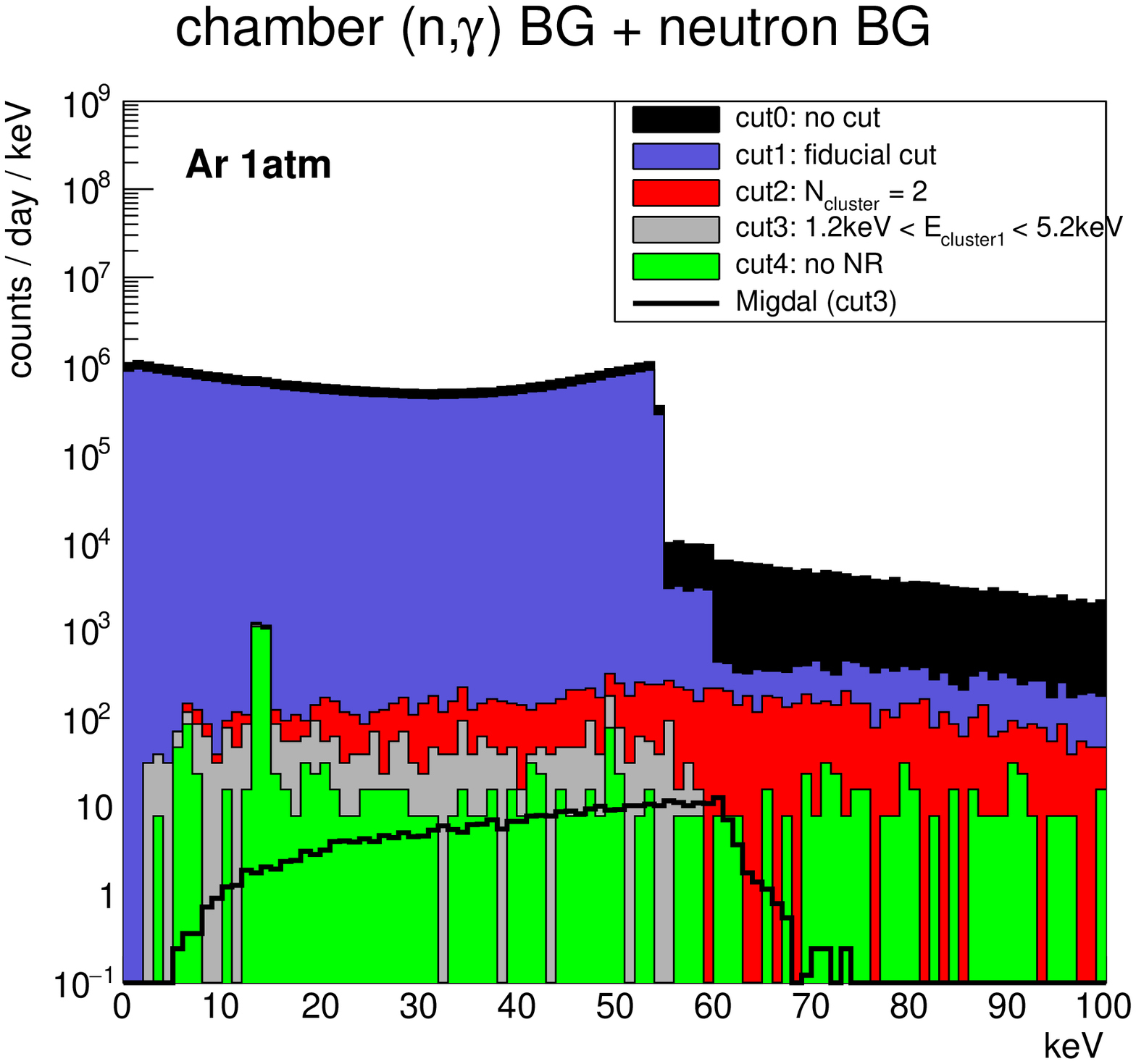} 
	    \caption{Results of the gamma-ray background MC simulation study. The energy spectra from the chamber material for the argon gas detector are shown. The colors are same as the ones used in Fig.~\ref{fig:Ar_neutron} with an additional green-filled histogram showing the gamma-ray background.} 
	    \label{fig:Ar_gamma_chamber} 
    \end{minipage}
    \begin{minipage}{0.49\hsize}
	    \centering
	    \includegraphics[width=0.99\linewidth]{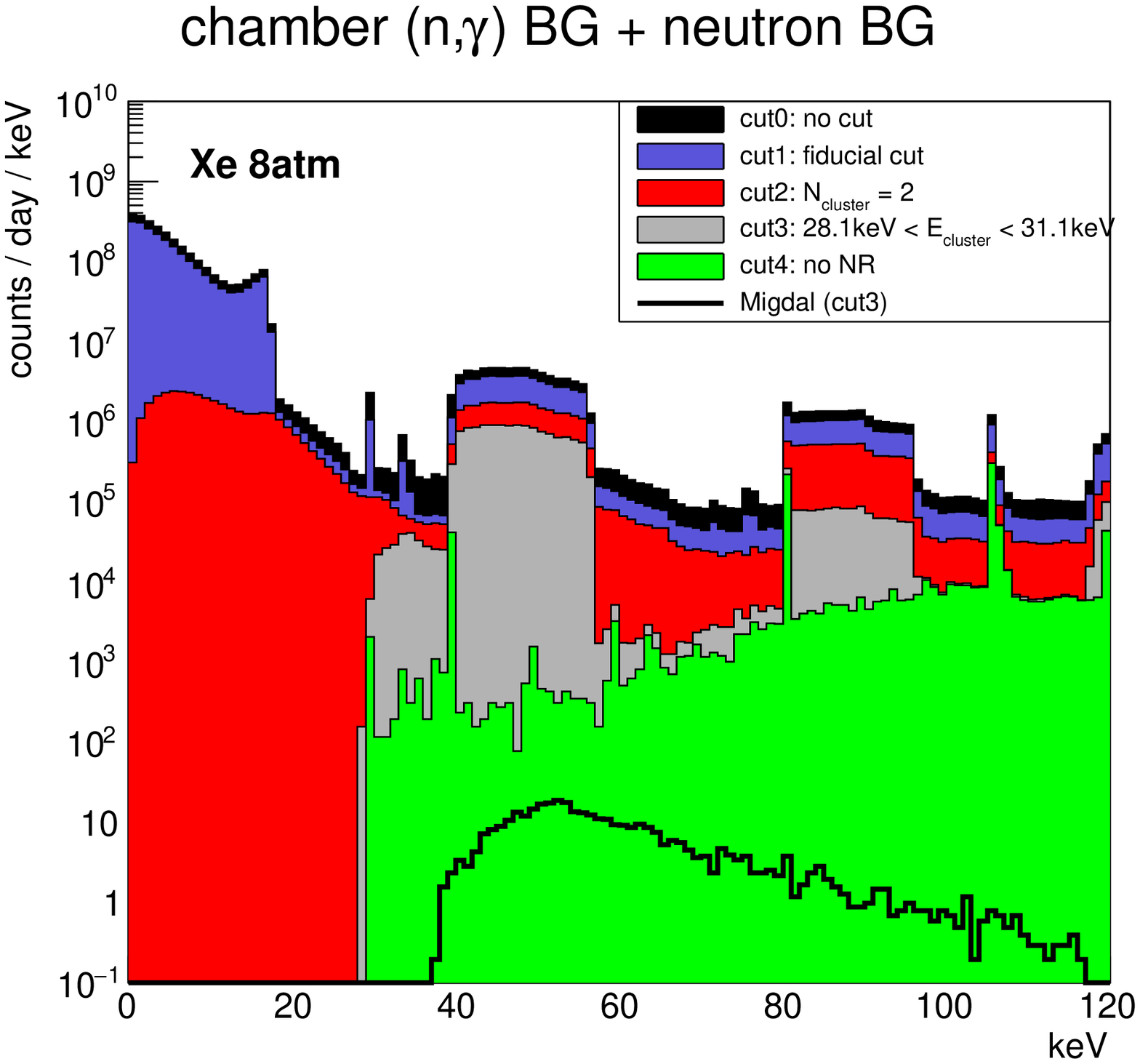} 
	    \caption{Results of the gamma-ray background MC simulation study. The energy spectra  from the chamber material for the xenon gas detector are shown. The colors are same as the ones used in Fig.~\ref{fig:Ar_gamma_chamber}.} 
	    \label{fig:Xe_gamma_chamber} 
	\end{minipage}
\end{figure}

\begin{figure}[htbp]
    \begin{minipage}{0.49\hsize}
	    \centering
	    \includegraphics[width=0.99\linewidth]{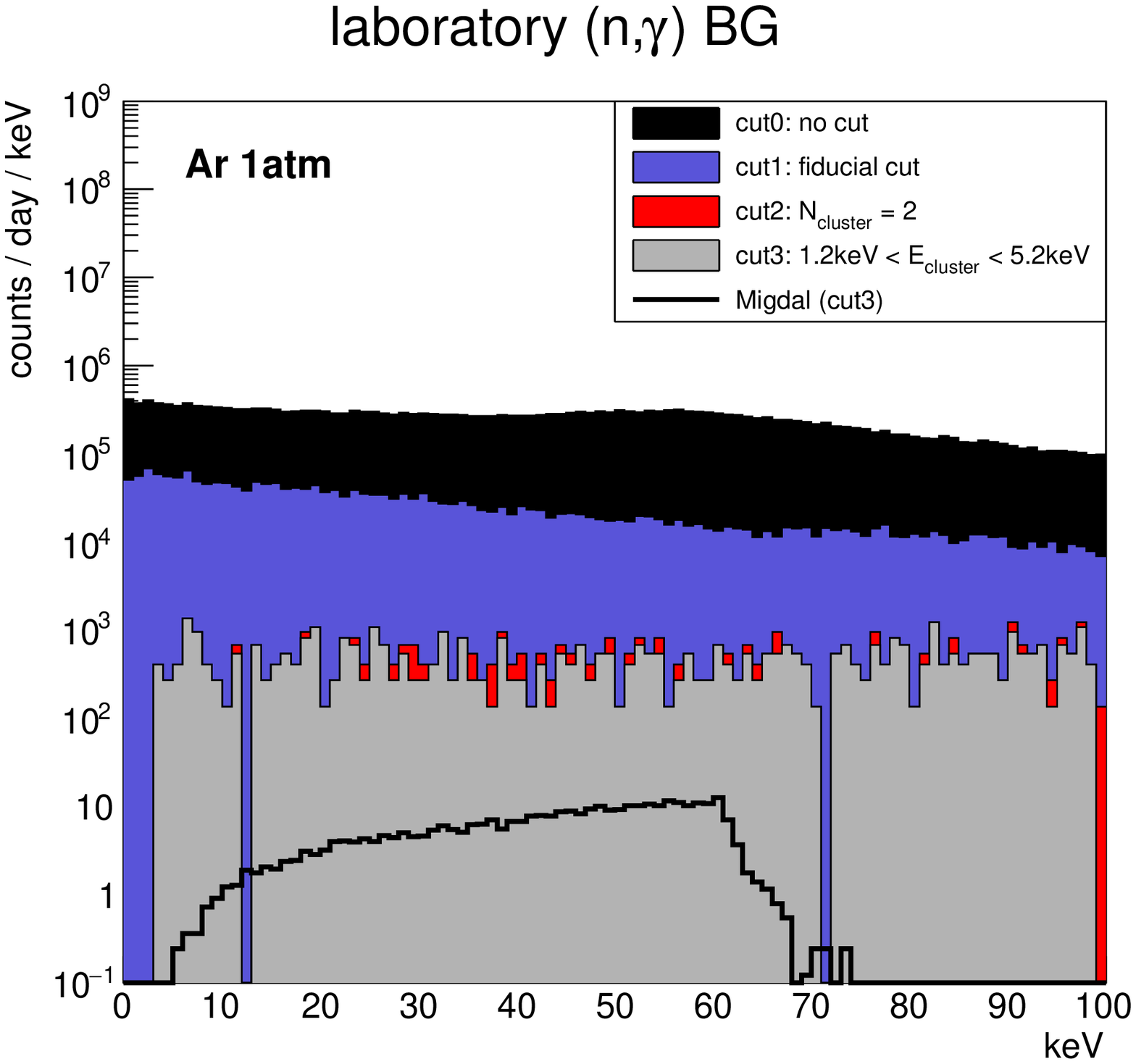}
	    \caption{Results of the gamma-ray background MC simulation study. The energy spectra from the laboratory material for the argon gas detector are shown. The colors are same as the ones used in Fig.~\ref{fig:Ar_neutron}} 
	    \label{fig:Ar_gamma_lab} 
    \end{minipage}
    \begin{minipage}{0.49\hsize}
	    \centering
	    \includegraphics[width=0.99\linewidth]{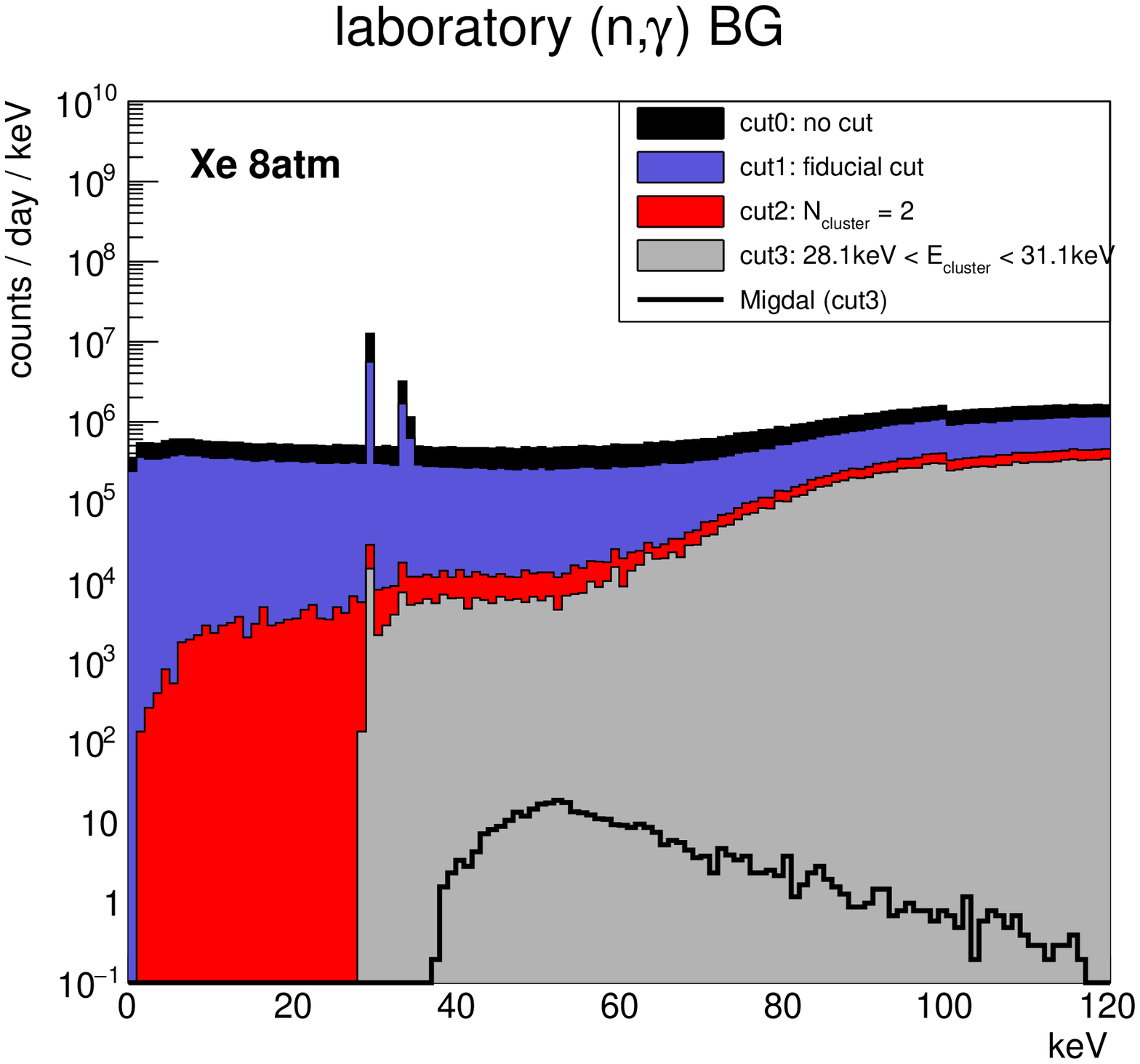} 
	    \caption{Results of the gamma-ray background MC simulation study. The energy spectra  from the laboratory material for the xenon gas detector are shown. The colors are same as the ones used in Fig.~\ref{fig:Xe_neutron}.} 
	    \label{fig:Xe_gamma_lab} 
    \end{minipage}
\end{figure}

The MC results are shown in Figs.~\ref{fig:Ar_gamma_chamber}$\sim$\ref{fig:Xe_gamma_lab}.
The selection criteria are same as the ones used in the intrinsic neutron backgrounds.
The gamma-ray backgrounds from the chamber material for the argon and xenon cases are shown in Fig.~\ref{fig:Ar_gamma_chamber} and Fig.~\ref{fig:Xe_gamma_chamber}, respectively.
The chamber has the gas medium inside so that these background spectra are the sum of the intrinsic neutron background and the gamma-ray background. 
The gamma-ray backgrounds from the chamber material are shown with green-filled histograms for a better understanding.
It is seen that the gamma-ray background rate is smaller than that of the intrinsic neutron background for the argon case. 
The gamma-ray background rate is also found to be 
smaller than that of the intrinsic neutron background but 
significantly larger than that of the Migdal signal for the xenon case.
Possibilities to reduce the chamber background are discussed in Section~\ref{sec:discussion}.

The gamma-ray background spectra from the laboratory material for the argon and xenon cases are shown in Fig. \ref{fig:Ar_gamma_lab} and Fig. \ref{fig:Xe_gamma_lab}, respectively.
(n,$\gamma$) reactions take places in the concrete and aluminum floor and the gamma-ray flux at the detector position is recorded. The gamma-rays are newly generated at the detector position to increase the statistics.
The rate of the gamma-ray backgrounds from the laboratory are found to be much larger than that of the Migdal effect. 
These backgrounds need to be reduced at the generation point of the neutrons. Hydrocarbon and $\rm^{10}B$ shiledings with a hole only to the detector direction can be set around the reaction point to reduce the number of neutrons reaching the laboratory materials. The gamma-rays from the laboratory can also be reduced with shieldings around the detector. It should be noted that these shieldings need to be designed with a good consideration of the laboratory geometry and neutron energy because these new material would be new sources of the gamma-ray background. With these shieldings, the gamma-ray background rates need to be reduced at least two orders of magnitude to observe the Migdal effect.

\subsection{Discussion}
\label{sec:discussion}
The background study so far assumes a continuous (DC) neutron beam. It would be possible to reduce delayed or continuous background 
such as laboratory gamma-rays with a pulsed neutron beam. 
The effect of the background rejection with the pulsed neutron beam would depend on the time evolution of the background rate.
One needs to have a good understanding of the neutron beam to estimate the background rejection by a pulsed-beam.
Gamma ray background from the chamber can be 
reduced by designing a chamber with a smaller amount of material like polyimide film for the argon gas case because the gas pressure is at the normal pressure.
For the xenon gas case, on the other hand, chamber material cannot be reduced because of its high pressure. Self-shielding technique can be used instead by increasing the detector volume and use outer $\sim$10~cm as a veto region. Also, it may be effective to install a solid scintillator on the wall inside the chamber \cite{SMILE2015} as an active shieldings.
In this sense this study is giving a conservative estimation of the background rate.

The angular distribution of the Migdal electron from $n=1$ follows $\cos^2\theta_{N-e}$, with $\theta_{N-e}$ as the polar angle from the momentum of the recoil nucleus.
Therefore, if the direction of the nucleus and the direction of the Migdal electron can be measured, the background can be reduced further.
This effect would be valid even by measuring only the direction of the Migdal electron because the direction of the recoil nuclei is biased in the forward direction.
The challenging point of this method is that the energy of the Migdal electrons is about 10~keV resulting a track length of 2~mm and 1~mm in the argon and xenon gas, respectively.
In the future, it would be interesting to be able to use the direction information of Migdal electrons by some breakthrough of the technology like emulsions, low-pressure gaseous detectors, or columnar recombination.

\section{Conclusions}
A possibility of experimental observation of the Migdal effect in the neutron scattering with position sensitive gaseous detectors is studied. 
Some modes of the Migdal effects would have two clusters, one is for the nuclear recoil and the other is for the associated X-ray.
Table-top sized position-sensitive gaseous detectors $\sim(30\rm cm )^3$ filled with argon or xenon target gas 
are found to 
be capable of detecting sufficient rates (O($10^2\sim10^3$) events/day) of these two-cluster Migdal events.
Distributions of the distance between two clusters, representing the absorption length of the characteristic X-rays,
 would be a strong evidence of the Migdal effect observation. 
Two significant background sources, namely the intrinsic neutrons and the neutron induced gamma-rays are found to exist.
These background rates are found to be higher than the Migdal effect rates even with a relatively low-energy (565~keV) neutron source.
It is found that the background rate can be reduced to the same order of the Migdal rate with a reasonable shielding against the gamma-rays from the laboratory for the argon case. 
The intrinsic background by inelastic scatterings would be a serious problem for the xenon case, which can be overcome with isotope separations. Remaining gamma-ray background from the chamber and the laboratory can be reduced with precisely-designed shieldings.
 As a consequence of this study, it is concluded that the experimental observation of the Migdal effect can be realized with a good understanding and reduction of the background.

\section*{Acknowledgment}
This work was supported by KAKENHI Grant-in-Aids (18K13567, 26104005, 16H02189, 19H05806, 18H03697, 18H05542, 19H05810), ICRR Joint-Usage,
I-CORE Program of the Israel Planning Budgeting Committee (grant No. 1937/12), World Premier International Research Center Initiative (WPI), MEXT, Japan.

\bibliographystyle{myptephy} %
\bibliography{bibfile}

\end{document}